\documentclass[12pt]{article}
\usepackage{amssymb,epsf}
\def\lsim{\mathrel{\rlap {\raise.45ex\hbox{$ < $}}
{\lower.45ex\hbox{$\sim$}}}}
\def\gsim{\mathrel{\rlap {\raise.45ex\hbox{$ > $}}
{\lower.45ex\hbox{$\sim$}}}}
\def\sqr#1#2{{\vcenter{\vbox{\hrule height.#2pt
        \hbox{\vrule width.#2pt height#1pt \kern#1pt
           \vrule width.#2pt}
        \hrule height.#2pt}}}}

\def\slsim{{\displaystyle
{{\raise-8pt\hbox{$\scriptstyle <$}}
\atop{\raise5pt\hbox{$\scriptstyle \sim$}}}}}
\def\sgsim{{\displaystyle
{{\raise-8pt\hbox{$\scriptstyle  >$}}
\atop{\raise5pt\hbox{$\scriptstyle \sim$}}}}}
\newskip\humongous \humongous=0pt plus 1000pt minus 1000pt

\newcommand{\sump}[0]{\sum_{(h,g)}\!{\raise 4pt \hbox{$'$}}\,}
\newcommand{\sumpf}[0]{\sum_{(H^{\rm f},G^{\rm f})}\!\!\!\!{\raise 4pt
\hbox{$'$}}\,}
\newcommand{\sumpo}[0]{\sum_{(H^1,G^1)}\!\!\!\!{\raise 4pt
\hbox{$'$}}\,}
\newcommand{\sumpF}[0]{\sum_{(H^F,G^F)}\!\!\!\!{\raise 4pt
\hbox{$'$}}\,}

\def\bs{\begin{subequations}}
\def\es{\end{subequations}}
\catcode`\@=11
\newcount\hour
\newcount\minute
\newtoks\amorpm
\hour=\time\divide\hour by60
\minute=\time{\multiply\hour by60 \global\advance\minute by-\hour}
\edef\standardtime{{\ifnum\hour<12 \global\amorpm={am}%
        \else\global\amorpm={pm}\advance\hour by-12 \fi
        \ifnum\hour=0 \hour=12 \fi
        \number\hour:\ifnum\minute<10 0\fi\number\minute\the\amorpm}}
\edef\militarytime{\number\hour:\ifnum\minute<10 0\fi\number\minute}
\def\draftlabel#1{{\@bsphack\if@filesw {\let\thepage\relax
   \xdef\@gtempa{\write\@auxout{\string
      \newlabel{#1}{{\@currentlabel}{\thepage}}}}}\@gtempa
   \if@nobreak \ifvmode\nobreak\fi\fi\fi\@esphack}
        \gdef\@eqnlabel{#1}}
\def\@eqnlabel{}
\def\@vacuum{}
\def\draftmarginnote#1{\marginpar{\raggedright\scriptsize\tt#1}}
\def\draft{\oddsidemargin -.2truein
        \def\@oddfoot{\sl preliminary draft \hfil
        \rm\thepage\hfil\sl\today\quad\militarytime}
        \let\@evenfoot\@oddfoot \overfullrule 3pt
        \let\label=\draftlabel
        \let\marginnote=\draftmarginnote
   \def\@eqnnum{(\theequation)\rlap{\kern\marginparsep\tt\@eqnlabel}%
\global\let\@eqnlabel\@vacuum}  }
\def\subequations{\refstepcounter{equation}%
  \edef\@savedequation{\the\c@equation}%
  \@stequation=\expandafter{\theequation}
  \edef\@savedtheequation{\the\@stequation}
  \edef\oldtheequation{\theequation}%
  \setcounter{equation}{0}%
  \def\theequation{\oldtheequation\alph{equation}}}

\def\endsubequations{\setcounter{equation}{\@savedequation}%
  \@stequation=\expandafter{\@savedtheequation}%
  \edef\theequation{\the\@stequation}\global\@ignoretrue
  \vspace*{-12pt} \\}
\def\bs{\begin{subequations}}
\def\es{\end{subequations}}
\relax
\def\thefootnote{\fnsymbol{footnote}}
\def\be{\begin{equation}}
\def\ee{\end{equation}}
\def\ba{\begin{eqnarray}}
\def\ea{\end{eqnarray}}

\def\Tr{\,{\rm Tr}\, }

\def\Im{\,{\rm Im}\, }

\def\D{{\cal D}}
\def\C{{\cal C}}
\def\su{SU(1,1)}
\def\ads{{{ADS}$_3$}}

\def\im{\, {\rm Im}\, \tau}
\def\e{{\rm e}}
\def\sp{\ , \ \ }

\def\nn{\nonumber}
\def\nl{\hfil\break}

\def\np#1#2#3{Nucl. Phys. {\bf{B#1}} (#2) #3}
\def\pl#1#2#3{Phys. Lett. {\bf{B#1}} (#2) #3}
\def\plo#1#2#3{Phys. Lett. {\bf{#1B}} (#2) #3}

\newcommand{\uarrw}[0]{\mathrel{
{\raise.5ex\vbox{\hrule width 1cm}\hskip-6pt\rightarrow}}}
\def\nl{\hfil\break}
\def\thebibliography#1{%
\vskip 0.5cm \centerline{\bf References}
\list{%
[\arabic{enumi}]}{\settowidth\labelwidth{[#1]}
\leftmargin\labelwidth
\advance\leftmargin\labelsep
\usecounter{enumi}}
\def\newblock{\hskip .11em plus .33em minus .07em}
\sloppy\clubpenalty4000\widowpenalty4000
\sfcode`\.=1000\relax}

\renewcommand{\theequation}{\arabic{section}.\arabic{equation}}

\renewcommand{\section}{\setcounter{equation}{0}\@startsection%
{section}{1}{0mm}{-\baselineskip}{0.5\baselineskip}%
{\normalfont\normalsize\bfseries}}

\renewcommand{\subsection}{\@startsection%
{subsection}{2}{0mm}{-\baselineskip}{0.5\baselineskip}%
{\normalfont\normalsize\slshape}}
\begin{document}
\renewcommand{\theequation}{\arabic{section}.\arabic{equation}}
\begin{titlepage}
\begin{flushright}
CPTH-PC-732.0899\\
hep-th/9908189\\
\end{flushright}
\begin{centering}
\vspace{.8in}
\boldmath
{\large {\bf STRING THEORY ON \ads :}}\\ 
\vspace{.1in}
{\large{\bf SOME OPEN QUESTIONS }}\\
\unboldmath
\vspace{.9in}
{P. Marios PETROPOULOS}\\
\vspace{.3in}   
{\it Centre de Physique Th\'eorique,
Ecole Polytechnique}$^{\ \diamond}$\\
{\it 91128 Palaiseau Cedex, France.}\\
\vspace{.9in}

{\bf Abstract}\\
\end{centering}
\vspace{.2in}
String theory on curved backgrounds has received much attention 
on account of both 
its own interest, and 
of its relation with gauge theories.
Despite the
progress made in various directions, several quite elementary questions
remain unanswered, in particular in the very simple case of
three-dimensional anti-de Sitter space.
I will very briefly review these problems, discuss in some detail the
important issue of constructing a consistent spectrum for a string
propagating on \ads\ plus torsion background, and comment on
potential solutions.
\vspace{0.9in}
\nl
{\sl To appear in the proceedings of the 
TMR European program meeting ``Quantum aspects of gauge theories,
supersymmetry and unification", Paris, France, 1--7 September, 1999.}
\vspace{.2in}
\begin{flushleft}
CPTH-PC-732.0899\\
August 1999\\
\end{flushleft}
\hrule width 6.7cm
\vskip.1mm{\small \small \small
$^\diamond$ Unit{\'e} mixte 7644 du {\it Centre National de la Recherche
Scientifique}.}
\end{titlepage}
\newpage
\setcounter{footnote}{0}
\renewcommand{\thefootnote}{\arabic{footnote}}

\boldmath
\section{String theory on \ads : the simplest setting beyond flat space}
\unboldmath

String theory is certainly the most appropriate setting for studying
quantum-gravity phenomena. This includes big-bang cosmology, black-hole
physics, and more general gravitational/gauge solitons or other exotic
objects.  In the absence of a truly non-perturbative approach to string
theory, the usual method consists in analysing the propagation of the
string on non-trivial backgrounds generated by some sources, which
correspond either to perturbative or to non-perturbative string states.
Consistency of string propagation imposes severe restrictions on the
allowed backgrounds, which must be conformal so as to satisfy the
whole set of requirements exactly in $\alpha '$. Approximations can also
be found by solving the relevant equations of motion up to some order
in $\alpha'$.

Three-dimensional anti-de Sitter (or de Sitter)
space was recognized long ago as a
case of interest with respect to the above motivations
\cite{dvs}--\cite{gaw}. It is a
maximally symmetric solution of Einstein's equations with negative
cosmological constant, and time is embedded non-trivially in the curved
geometry. Alternatively, it corresponds to the Freedman--Gibbons
electrovac
solution of gauged supergravity, which can be shown to leave space-time
supersymmetry unbroken \cite{abs}. Other peculiar features of \ads\ 
are the absence of
asymptotically flat regions, the presence of boundaries
(when conformally compactified), as well as a rich
causal structure, which makes it possible to obtain three-dimensional
black holes after modding out some discrete symmetry \cite{btz}.

As far as string theory is concerned, \ads\ 
is an {\it exact} background, provided
an NS--NS two-form is switched on. In fact, three-dimensional anti-de
Sitter space is the universal covering of the $\su $ group manifold, and
the corresponding two-dimensional conformally invariant sigma model is a
Wess--Zumino--Witten model, which naturally accounts for the torsion
background.

Several remarks can be made here in order to argue that \ads\ 
provides the
simplest setting for string theory beyond flat space. General non-compact
group manifolds define a natural framework for studying strings on
space-times with non-trivial geometry. Restricting ourselves to the case
of simple groups, however, only $\su $ possesses a single time direction;
\ads\ 
is therefore {\it the only} exact background where string propagation
leads
to a WZW model. Of course, cosets with one time direction can be
constructed out of simple non-compact group manifolds. This is the case,
for instance, 
for ADS$_n$, which appears as 
$O(2,n-1)/O(1,n-1)$.
However, these geometries cannot be obtained by the usual GKO construction
in
the framework of gauged WZW models. 

Last but not least, the motivation for understanding string theory on
\ads, and
more generally on ADS$_n$, is related to the recent developments on
ADS/CFT correspondence~\cite{adscft}. There, supergravity in the bulk of
anti-de Sitter
space is argued to be in some sense equivalent to a large-$N$
super-Yang--Mills theory on the boundary. Since the supergravity theory under
consideration is the low-energy limit of a more fundamental superstring
theory, the question arises of the exact structure
of the latter on the anti-de Sitter background, and its connexion to the 
super-Yang--Mills theory on the
boundary. Here also \ads\ 
plays a particular role. The
asymptotic isometry group is infinite-dimensional
\cite{bh}, and the theory on the
boundary is a two-dimensional conformal field theory. The latter is
different from the two-dimensional sigma model whose target space is the
``bulk \ads" on which the string propagates. Considerable efforts have
been made for understanding the relationship between these two
conformal theories, in order to both set more precisely the ADS/CFT 
correspondence, and try to get some feedback on the structure of the
string theory on {\ads}~\cite{boundconf}\footnote{Notice that most of
these works deal with ``Euclidean \ads", $H_+^3$, whereas I will present
here the 
ordinary Minkowskian situation. Except for the unitarity properties, the 
two cases share many features.}. 

Despite those efforts and the apparent simplicity of the model at hand, I
will show that
several important and elementary issues, such as the determination of the
spectrum,
consistent with the basic requirements of string theory
and conformal field theory, are still beyond our understanding. I will
also try to motivate various suggestions for further analysis.

Although we are ultimately interested in understanding how
superstrings behave on \ads\ background, I will concentrate in the
sequel on the bosonic case, where the issues I would like to address
are already visible. Moreover, this case might have some relevance in
the framework of recent attempts at establishing some relationship
between various bosonic theories -- including perhaps the celebrated
$26$-dimensional theory.

\boldmath
\section{The $SU(1,1)$ Wess--Zumino--Witten model}
\unboldmath

The analysis of string theory on \ads\ plus torsion background   
can be performed in two steps. First, we must study the sigma model 
whose target space has the above geometry; this is a WZW model on
the $\su$ group manifold. Then, the latter has to be coupled to
two-dimensional gravity. At the level of the Hilbert space, this
amounts to the decoupling of a certain subspace, which becomes
unphysical.

As a general remark, it should be stressed here that the geometrical
interpretation of a conformal field theory as a string propagating in some
backgrounds, is sometimes loose. It becomes unambiguous only in some
semi-classical limits, or in the presence of a dense spectrum of
Kaluza--Klein modes. Hence, one should be aware that very often one
is not describing the situation for which the model was designed.
Conversely, unexpected geometrical interpretations may arise.

Very little is known about WZW models on non-compact groups, at a
sufficiently rigorous and general level. Most of our knowledge is based
on a formal extension of the compact case to some specific situations,
and in the framework of current-algebra techniques.
Target-space boundary conditions, in particular, are treated somehow carelessly,
although we know how important they are for selecting various
representations when studying quantum mechanics on
{\ads}~\cite{oldads3, bs}, or in
the
determination of the asymptotic symmetry algebra acting on that 
space~\cite{bh}. This should be
kept in mind in any attempt to go beyond our present knowledge of the
subject. 

We usually assume that the $SO(2,2)\cong \su_{\rm L} \times \su_{\rm R}$
symmetry of the above model is realized in terms of an affine Lie
algebra, the level of which is not quantized because $\pi_3(\su)=0$
or, put differently, because of the absence of any Dirac-like singularity
in the torsion background.

The commutation relations for the modes of
the currents ($J^{a}_{\vphantom a}(z)=\sum_{m\in {\bf Z}}
z^{-m-1}J_{m}^{a}$, $a=1,2,3$) are
$$
\left[J_{m}^{a},J_{n}^{b}\right]=
if_{\; \; \; c}^{ab}\, J_{m+n}^{c}+{k\over 2}mg^{ab}_{\vphantom a}\delta
_{m+n}\; ,
$$
with
$g^{ab}_{\vphantom a}={1\over 2}f_{\; \; \; c}^{bd}\, f_{\; \; \; d}^{ca}=
{\rm diag}(--+)$
and $f_{\; \; \; d}^{ab}\, g^{dc}_{\vphantom a}=
-\epsilon^{abc}_{\vphantom a}$
($\epsilon^{123}_{\vphantom a}=1$). We expect the anomaly $k$ to be
negative so that
there is only one negative-metric generator that plays the role of 
the time coordinate, namely the third direction. We also introduce
$J_{m}^{\pm}=iJ_{m}^{1}\mp J_{m}^{2}$.

The world-sheet energy--momentum tensor is 
given by the affine Sugawara construction:
$$
T(z)={1\over k+2}\, g_{ab}:J^{a}(z)\, J^{b}(z): \; .
$$  
The modes $L_m$ 
($T(z)=\sum_{m\in {\bf Z}} z^{-m-2}L_{m}$)
satisfy the Virasoro algebra
with central charge
$c={3k\over k+2}$, and 
\be
\left[L_{m}^{\vphantom a},J_{n}^{a}\right]=-n
J_{m+n}^{a}\; .
\label{com}
\ee

Finally, the Hilbert space is formally constructed as in the compact  
case: it is a direct sum of products of representations of the left   
and right current algebras.
Highest-weight representations\footnote{Representations without highest
or lowest weight do exist \cite{nhw}. It is, however, not
clear how those could be interpreted within a stable string theory.
More representations of the $\su$ current algebra can be found
in \cite{recur}.}
of
the $\su$ current algebra are labelled by the spin $j$ of the primary
fields (states of level zero), which form a representation of the
global algebra (generated by the zero modes $J^{\pm , 3}_0$), and have
conformal weight $j(j+1)/k+2$.

Irreducible representations of the global algebra are essentially of two
kinds 
\cite{math}: discrete $\D^{\mp}(j)$ or continuous principal
$\C_{\rm p}(b,a)$ 
and continuous supplementary
$\C_{\rm s}(j,a)$. The discrete ones have highest ($\D^-$) or lowest
($\D^+$)
weight, whereas the continuous ones do not. The spin $j$ of the discrete
representations is real\footnote{In order to avoid closed time-like
curves, we are considering the universal covering of $\su$.
Therefore, $j$ is not quantized.}, and their states are labelled by
$\vert j m\rangle$, $m=j, j\mp 1, j \mp 2, \ldots $ For the
principal continuous ones, $j=-{1\over 2}+i b,\ b<0$, and the magnetic
number is $m=a, a\pm 1, a\pm 2, \ldots  ,\ -{1\over 2} \leq a<
{1\over 2}$, $a,b \in {\bf R}$; for the supplementary continuous
ones, $-{1\over 2}
\leq j<0$ and
$-{1\over 2} \leq a< {1\over 2}$, with the constraint
$\left\vert j + {1\over 2}\right\vert < {1\over 2}-\vert  a \vert $,
$a,j \in {\bf R}$. 
These representations are unitary and infinite-dimensional;
$\D^{\pm}(j)$ become finite-dimensional when $j$ is  a positive integer
or half-integer, and are non-unitary for any positive $j$.
Notice finally that the quadratic Casimir ($j(j+1)$) is negative for both
continuous series; for the discrete ones it is negative or positive
when $-1<j<0$ or $j<-1$, respectively.

Highest-weight representations of the current algebra are obtained
by acting with $J^{\pm , 3}_{-1}$ on the above level-zero
states, which are annihilated by all positive-frequency modes. These
representations contain an infinite tower of negative-norm states,
due to the indefinite metric $g^{ab}$. Therefore, in the above
setting, it is impossible to write down a unitary conformal theory
based on the $\su$ WZW model. This is not surprising, and the same
conclusion holds anyway for three free bosons with metric $(-++)$,
obtained here when $k\to -\infty$.

I now come to the following crucial question: How should the above   
representations be combined to form a consistent, though non-unitary,
model? In order to answer this question, we can follow the requirement
of {\it modular invariance}. The genus-one partition function reads, in   
general:
\be
Z(\tau,\bar \tau) = \sum_{{\rm L},{\rm R}} {\cal N}_{{\rm L},{\rm R}}\, 
\chi_{\rm L}(\tau)\,
\bar \chi_{\rm R}(\bar \tau)\; ,
\label {zgen}
\ee
where the summation is performed over all left--right representations
present in the spectrum, and $\chi(\tau)$ are the corresponding
characters:
\be
\chi(\tau)= \Tr q^{L_0 - {c\over 24}}\; ,
\label{ch}
\ee
$q= \exp 2i\pi\tau$. The multiplicities ${\cal N}_{{\rm L},{\rm R}}$
must be chosen in such a way that the partition function is invariant
under $\tau \to \tau +1$ and $\tau \to -1/ \tau$. Notice that, again,
Eq. (\ref{zgen})
is formal in the non-compact case, and one should prove it, e.g.   
by using path-integral techniques starting directly from the WZW
action, as in Ref. \cite{gaw}.

Already at this level, a major problem appears, which is actually
generic to all non-compact groups. The unitary
representations of
the global algebra being infinite-dimensional, there is an infinite
degeneracy level by level in the representations of the current
algebra, and consequently the characters (\ref{ch}) are 
ill-defined\footnote{We could consider finite-dimensional
non-unitary representations of the global algebra, since the Verma
module built on any representation is anyway non-unitary. However,
for later use in string theory, this choice would not be sensible.}.
This is the price to pay for using the full {\it non-compact and
non-Abelian symmetry} to   
classify the states of the theory. In fact, this is not specific to
the two-dimensional sigma model we are analysing. Similar problems
would occur in the relativistic quantum mechanics of a particle on a
two-dimensional plane, if we 
tried to describe its propagator by using wave-function representations
of the full Lorentz group $SO(2,1)$. The reason why free bosons can be
analysed without trouble just relies on the Abelian nature of the
symmetry used to classify their spectrum.

In our formal treatment, the only way out is to lift the degeneracy by
switching on a source coupled to $J_0^3$:
\be
\chi (\tau, v)= \Tr q^{L_0 - {c\over 24}}
\,
\e^{2i\pi v J_0^3}
\; .
\label{chr}
\ee
Notice, however, that this definition {\it does not} allow a
regularization of
the characters of the representations based on the continuous
series\footnote{For those, one could replace $J_0^3$ by
$\vert J_0^3\vert$ in Eq.~(\ref{chr}). Such characters have never been
studied in the mathematical literature.}. This shows that discrete and
continuous representations definitely play different roles, and that the
continuous ones do not fit into the present current-algebra approach. 
Moreover, convergence of the trace in (\ref{chr}) demands $\Im v >0$ for
$\D^+$ and  $\Im v <0$ for $\D^-$. As a consequence, within the present
framework,  $\D^+$ and  $\D^-$  {\it cannot appear simultaneously in the
spectrum.} 

In computing the characters, the main difficulty is to properly
identify the {\it singular vectors}.  These are zero-norm states   
orthogonal to any other state; their descendents possess the same  
property and they are thus responsible for the reducibility
of the Verma modules. Exhaustive and rigorous results can be found in
Refs. \cite{mff}. 

There are some particular sets of
representations of the current algebra: the {\it admissible}
representations\footnote{In the context of the $SU(2)$ WZW,
using the GKO coset construction, these series lead to the minimal   
BPZ models with $c<1$. For the integer level ($u=1$ in Eq.~(\ref{levad})),
unitarity is guaranteed, and one gets the $ADE$ invariants for
$SU(2)$ as well as the corresponding unitary series at $c<1$.}. These
are based on the discrete series, and appear at the level
\be
k={t\over u}-2 \sp t\ge 2 \sp u>0 \sp t,u \in {\bf Z}\; ,
\label{levad}
\ee
with spins
\be
j={1\over 2}\left(n-s\,{t\over u}\right)\sp
0\leq n\leq t-2 \sp
0\leq s\leq u-1 \sp n,s \in {\bf Z}\; .
\ee
For these representations, $k\ge -2$, and the spin obeys the following
bounds:  
$$
{1-u\over u}\, {t \over 2}\leq j \leq {t \over 2}-1\; .
$$
The primary states do not necessarily belong to some
unitary representation of the global algebra, since $j$ can be
positive. Singular vectors appear at various levels, which makes
the evaluation of (\ref{chr}) quite intricate.
Nevertheless, the characters of these series were obtained
in \cite{kw}. They turn out to form finite representations of the
modular group, which now acts as: 
$$
T\, :\; (\tau ,v )\to (\tau +1,v ) \sp
S\, :\; (\tau ,v )\to \left(-{1\over \tau },
{v \over \tau }\right)   \; .
$$

Rational models with an $ADE$ type of classification can be
constructed by using the above results \cite{cs}. Besides being
non-unitary, these models have peculiar properties. Their central
charge
is given by $c=3-6u/t$, which is negative when $t<2u$. Moreover, the
above models are only defined in the presence of a ``magnetic field",
which is not invariant under modular transformations. The
interpretation of these features is not clear.

For string-theory purposes, the level of the current algebra should
satisfy $k<-2$: this ensures positive central charge as well as a
single time-like direction in the target space, which are both
necessary conditions for the physical spectrum to be free of
negative-norm states. Furthermore, as far as the discrete series are
concerned, $j$ should be non-positive in order to avoid unitarity
problems already at level zero. This excludes the admissible
representations, and therefore the possibility of using their
modular-invariant combinations.  
  
In the regime $k<-2$, very little is known about the characters
of the $\su$ current algebra.
Those characters can be computed in the case
of highest-weight representations based on discrete series
(Eq. (\ref{chr})), for generic
values of $k$ and $j$, where {\it no singular vectors} are present, with
the result \cite{bk, tm}:
\be
\chi _{j}^{k}(\tau ,v)=
{q^{(2j+1)^{2}\over 4(k+2)}\, \e^{{i\pi\over 2}(2v+1)(2j+1)}
\, \e^{-i\pi j}
\over  \vartheta _{1}(\tau ,v ) } \; ,
\label{chg}
\ee
where $\vartheta_1(\tau,v)$ is the odd Jacobi function   
$$
\vartheta_1(\tau,v)=-2\sin (\pi v) \, q^{1\over 8}
\prod_{n=1}^{\infty}\left(1-q^n\right)
\left(1-q^n\, \e^{2i\pi v}\right)
\left(1-q^n\, \e^{-2i\pi v}\right)\; .
$$

Expression (\ref{chg}) does not hold for some discrete sets
of $(k,j)$'s, such as when $2j-k+n=1$, $n$ a positive integer, or
when $j=m/2$, $m\in {\bf Z}$, independently of $k$. In such cases, the
presence of null states will obviously spoil (\ref{chg}). In those
situations it is probably more of a technical problem than a
conceptual one to determine the exact characters. A much more
difficult issue is certainly how to combine the various characters
for obtaining modular-invariant partition functions. As an example,
we can consider the modular transformations of the characters
(\ref{chg}). We obtain:
\ba
\chi _{j}^{k}( \tau +1,v )&=&
\e^{{i\pi \over 2}\left({(2j+1)^{2}\over (k+2)}-{1\over 2}\right)}\,
\chi _{j}^{k}( \tau ,v )
\; , \label{mt1} \\
\chi _{j}^{k}\left(-{1\over \tau },
{v \over \tau }\right)&=&
\sqrt{2\over k+2} \, \e^{{i\pi \over 2}\left({v^2\over \tau}k + 1\right)}
\int_{-\infty}^{+\infty} {\rm d}{\ell} \,
\e^{-i\pi{(2j+1)(2\ell +1)\over k+2}}
\,
\chi ^k_{\ell}(\tau, v)\; .\label{mt2}  
\ea
These transformations involve {\it all} values of $j$,
with
zero measure for
the  discrete sets of representations possessing singular vectors.   
Constructions involving only these generic characters turn out to be 
too simple, and do not enable us to obtain interesting modular-invariant
combinations. In particular, the naive diagonal 
combination integrated over all values of $j$
(again, all primary states do not belong to
unitary representations of the global algebra) is not,
strictly speaking, modular-invariant
because of the $v$-dependent prefactor appearing in (\ref{mt2}).
Nevertheless,
considering this combination, we find:
\be
Z_{\rm diag}^k (\tau, v)= \int_{-\infty}^{+\infty} {\rm d}j 
\left\vert
\chi ^k_j(\tau, v)
\right\vert^2
={1\over 2}\, \sqrt{k+2\over \im}\,
{\e^{\pi(k+2){(\Im v)^2\over\im}}
\over
\left\vert
\vartheta_1(\tau,v)
\right\vert^2}
\label{dni}
\ee
(as usually, in the presence of a time-like coordinate, analytic
continuation is needed -- here when $k<-2$ in Eqs. (\ref{mt2}) and
(\ref{dni})). Under an $S$-transformation,
an extra factor appears: $\left\vert 
\exp i\pi k v^2/\tau
\right\vert$. The latter is irrelevant at $v \to 0$. Since this limit is
singular, however, modular invariance should be demanded for any finite
value of $v$. It can be reached only if, in expression (\ref{dni}), 
the measure ${\rm d}j$ is replaced with ${\rm d}j\, \exp - \pi k{(\Im
v)^2\over\im}$. This formally defines an invariant combination at any
$v\neq 0$, because it accounts for the cancellation of the extra
$v$-dependent factor appearing in the transformation (\ref{mt2}). However,
in this way, the diagonal combination no longer depends on $k$ (except for
the overall volume factor $\sqrt{k+2}$), which means in particular that
{\it
the asymptotic behaviour of the spectrum does not depend on the central
charge.} This situation is hardly acceptable (another argument is given
at the end of Section 3 for the string theory), and the above results
should be interpreted as a
sign that, among others, we should consider more carefully the appearance
of representations with singular vectors. I will come back to this point
when studying the string on \ads.
It is also interesting to observe that the result (\ref{dni}) was
obtained in \cite{gaw} as the  {\it partition function of Euclidean
three-dimensional anti-de Sitter space, $H_+^3\equiv SL(2,{\bf
C})/SU(2)$,}
by using a rigorous path-integral approach\footnote{See also \cite{tes}
for a rigorous treatment of  $H_+^3$. In
Ref. \cite{gaw}, the
two-dimensional Euclidean black hole was also analysed. For the latter,
the result turns out
to be
modular-invariant.}. It is not clear why such a non-modular-invariant
partition function would be satisfactory in the case of $H_+^3$. 

So far, I have been considering the construction of conformal models
based on $\su$ WZW at level $k$. The encountered problems can be
summarized as follows. One is the infinite degeneracy at each level
in the representations of the current algebra, and in particular the
treatment of the representations based on the continuous series for
which no character formula has been proposed in the mathematical
literature. 
This problem might be fixed in a
path-integral approach, where a zero mode responsible for the
corresponding 
(infinite-volume-like)
divergence could be identified and removed. Alternatively, 
we might also need a
deformation of the affine Sugawara construction in order to lift the
degeneracy without coupling to an external field. Modification of the
current algebra itself has also been advocated \cite{bars}. The
question then arises whether these deformations still describe the
initial WZW theory. For example, in the compact-group WZW models, the
natural stress tensor, obtained by differentiating the action with respect
to the metric, is precisely the one given by the affine Sugawara
construction \cite{hal}. Any deformation with respect to the latter,
possibly continuous and conformal, will abandon the original WZW theory.

Another problem is related to the construction of various
modular-invariant partition functions: What are the sets of
representations -- including representations based on both 
discrete and continuous series -- which form a well-behaved OPA? Only
the sets of admissible representations, based on some discrete series,
have been identified.

This question is difficult and we can somehow understand why by
comparing our case to the situation of a WZW model on the group
manifold of $SU(2)$. The $SU(2)$  theory can be unitary because the
affine
algebra has unitary highest-weight representations for integer and
half-integer spin such that $0\leq j\leq k/2$ ($k$ is integer here).
Modular invariance is therefore expected for combinations of
representations falling within this range, and indeed this happens.
That is not a miracle: the structure of characters, and thereby their
modular transformations, is directly dictated by the presence of    
singular vectors, which in turn determine the unitarity properties  
of the representations since they appear as limiting cases of
positive-norm states becoming negative-norm. For example, the
(sufficient) condition $2j - k +n=1$ for having a null state at
level $n$ has solutions within the range $j\leq k/2$,
which embeds the unitarity domain. Another
instructive example is the case of the free boson. There, all
representations of the $U(1)$ algebra are unitary -- none if the boson
is of time-like signature -- and are labelled by a continuous
momentum. No null states appear and all representations must be used
in a consistent model. They lead to the celebrated $\left(\sqrt{\im} \eta
\bar \eta\right)^{-1}$ partition function. Both for the $SU(2)$ WZW model
and for the free boson, unitarity is a guideline for reaching
modular-invariance.
For $\su$ there are {\it no}
unitary highest-weight representations of the current algebra,
whereas some have null states and some others do not. Unitarity
and presence or absence of singular vectors
cannot therefore be successfully advocated for constructing
modular-invariant combinations.

\boldmath
\section{String theory on $SU(1,1)$}
\unboldmath  

I will now analyse the string propagating over the $\su$
group manifold. The coupling of the above conformal model to the
two-dimensional gravity creates spurious states that we should eliminate
from the spectrum. The most straightforward approach would have been the
light-cone-gauge analysis. Unfortunately, this method is hard to
implement (despite several attempts \cite{lc}) and we have
therefore to advocate -- without rigorous proof -- that going to the
conformal gauge and imposing the Virasoro constraints will eliminate all
spurious states, provided the conformal anomaly is cancelled. 

In this analysis, the natural questions are the following: What are the
representations of the  $\su_{\rm L} \times \su_{\rm R}$ current algebra
that should be kept in order to have a consistent theory
(in particular a well-behaved operator algebra)? 
What are the roles of discrete versus continuous representations?
Are all physical
states of positive norm? What kind of particles do these states 
describe, and what are the corresponding vertex operators?

Anomaly cancellation for the bosonic string implies 
$c=26$,
where $c$ accounts for all matter-field contributions. String theory on 
$\su$ can be critical on its own, since the level of the affine algebra
can be freely tuned to reach the critical central charge: $k=-52/23$. It
might be
relevant, however, to keep $k$ free, and couple the $\su$ sigma model to
some unitary conformal field theory such as $d$ free bosons, a WZW model
on $SU(2)$, \dots \
This can help in understanding the theory at large
$\vert k \vert$, corresponding to the near-flat-space limit. 

As was emphasized in the previous section, very little is known about the
WZW model on $\su$. The only guideline is therefore the search for
representations of the current algebra leading to a positive-definite
physical Hilbert space. In fact, it is straightforward to argue that
within
the class of highest-weight representations of the $\su$ current algebra
we have considered, and in the general framework we have presented so far
for analysing the string propagation on a non-compact manifold, {\it
there is no satisfactory selection} of representations that can be
performed, which guarantees the absence of negative-norm states in the
physical Hilbert space.

The argument goes as follows. I will concentrate on the left-movers,
keeping in mind that they should be eventually paired with right-movers.
A highest-weight representation of the
current algebra is built on a representation of the global algebra, which
defines the level-zero states and is annihilated by positive-frequency
current modes. Acting on those states with $J^{\pm, 3}_{-1}$ will generate
the Verma module. At each level, the set of states can be decomposed with
respect to the global algebra. Since the Virasoro generators commute 
with the modes $J^{\pm, 3}_0$ (see Eq.~(\ref{com})), Virasoro constraints 
($L_m\vert {\rm physical}\rangle=0 \ \forall m>0$) will keep or throw
away
complete representations of the global algebra. This 
considerably simplifies the rules for implementing unitarity:
(\romannumeral1) the
level-zero states should all have positive norm, i.e. be a unitary
representation of the global algebra of the type $\D^{\pm}(j)$, $\C_{\rm
p}(b,a) $ or $\C_{\rm  s}(j,a)$
(see previous section for the allowed values of the parameters $j,a,b$);
(\romannumeral2) at each level, any physical representation should 
also have parameters consistent with unitarity; this last statement
ensures that all states of the representation at hand are positive-norm,
provided the norm of one of them is indeed positive.

A simple computation shows that, irrespectively of the type of unitary
level-zero representation, $\D^{\pm}(j)$, $\C_{\rm
p}(b,a) $ or $\C_{\rm  s}(j,a)$, at level one there will be generically
three representations of the global $\su$ algebra: two Virasoro primaries
(i.e. physical up to mass-shell condition) with spin $j+1$ and $j-1$, and
an unphysical one with spin $j$. This generalizes at level $N$, where we
meet
at least two Virasoro-primary representations, with spin $j\pm N$.

In the case of continuous series $\C_{\rm p}(b,a) $ or 
$\C_{\rm  s}(j,a)$, already at level one, the values of the
spin are out of the unitarity range: for $\C_{\rm p}(b,a) $ the quadratic
Casimir becomes complex, whereas it becomes positive for 
$\C_{\rm s}(j,a)$. On the other hand, at level $N$, mass-shell condition
reads:
\be
{j(j+1) \over k+2} + N \leq 1
\label{msc}
\ee
(an internal positive conformal weight is supposed to compensate the
difference
with respect to $1$); this
shows that the maximal allowed level for a primary spin-$j$
representation is 
\be
N_{\rm m}(j)={\rm integer \ part}\left(
1-{j(j+1) \over k+2}
\right)\; .
\label{ml} 
\ee
Therefore, as far as the continuous series are concerned, in the regime of
interest ($k< -2$), $ N_{\rm m}(j)=0$. Unitarity is guaranteed, but 
these representations only describe part of the {\it tachyonic sector} of
the
theory \cite{tm}.

The case of discrete series goes along the same lines. For $k< -2$,
all states have positive norm in the range $-1<j<0$, but are all tachyonic
(level zero only is allowed). When $j\leq -1$, $ N_{\rm
m}(j)\geq 1$, which corresponds to more general massive, massless or
tachyonic
excitations. However, since $ N_{\rm m}(j)$ grows quadratically with $j$,
for sufficiently large $\vert j \vert$, $ N_{\rm m}(j) + j$ becomes
positive, and physical non-unitary representations appear
\cite{bfow}. Negative-norm
states remain in the spectrum\footnote{As expected, the situation for 
$k> -2$ is worse and unitarity is definitely lost in that case.
For the continuous series, $ N_{\rm m}(j)\geq 1$. Thus, for any spin,
non-unitary physical representations appear at several levels. The same
conclusion holds for the discrete series with $-1\leq j <0$, whereas
only tachyons are physical for $j<-1$.}.  

The situation described above is not very encouraging. Representations of
the current algebra based both on continuous and discrete series seem to
be required for generating the complete bosonic spectrum. Virasoro
constraints and mass-shell condition guarantee unitarity for the
continuous series
-- which describe only tachyons --, 
but do not succeed in the case of discrete
representations. When the spin is of order $j\lsim k+2$, the norm of the
on-shell states in the current-algebra representation is no longer
positive; for these states $M^2\gsim \vert k+3 \vert$.

Within this framework, if we insist on having a theory free of
negative-norm states, the only possibility left would be to cut the spin
$j$ ($j_{\rm min}\leq j <0$) in such a way that 
$N_{\rm m}(j_{\rm min}) + j_{\rm min}$ be non-positive \cite{pmp}. We thus
avoid some
physical non-unitary representations that would have been present 
otherwise, and there is hope that all negative-norm states decouple 
in this way.

It is important to be aware that {\it this latter possibility violates 
the generic structure of the string spectrum itself}. We lose the
infinite
tower of string modes (the mass is cut off at the scale of the
radius of \ads), and consequently the hope of
constructing a consistent spectrum shrinks. Modular invariance is 
expected to be spoiled.
Moreover, we cannot even keep the unit representation in the spectrum,
namely the current-algebra representation with  $j=0$, since 
$N_{\rm m}(j=0)=1$, and level one contains a representation of the global 
algebra with spin 1, which is not unitary.
Despite these features the above possibility has been worked out because
of some appealing properties. Let me briefly summarize the 
situation.

Starting from a level-zero unitary discrete representation $\D^-(j)$
($j<0$), we find a representation $\D^-(j+1)$ at level one, which is
Virasoro-primary. The norm of its highest weight
is given by $2j-k$. The mass-shell condition implies that this
representation 
is present as
long as $j\leq -1$ (remember $k<-2$). In that range, unitarity thus 
demands
\be
{k\over 2}\leq j<0\; .
\label{uc}
\ee

Condition (\ref{uc}) is the key of our analysis \cite{pmp, moh}. It is
similar to the condition appearing in $SU(2)_k$ and has in fact the same
origin, although its purpose here is not to garantee the unitarity of the
$\su$ WZW model, but the unitarity of the latter modded out by the
Virasoro constraints. The above condition on the spin has drastic
consequences over the string spectrum. By using Eq.~(\ref{ml}), there 
appears
an {\it absolute upper bound on the string level},
\be
N_{\rm max}={\rm integer \ part}\left(1-{k \over 4}\right)\; ,
\label{mml}
\ee
and similarly for the mass squared. For instance, if the string is a pure
WZW model on $\su$, $k=-52/23$, and the physical spectrum is made out of
tachyons and massless states only. On the other hand, we can add an 
internal unitary conformal field theory with positive central charge
$c_{\rm int}$. The bigger $c_{\rm int}$ is, the larger  $\vert k
\vert$ is,
and more and more massive are the
states that
the physical spectrum 
acquires\footnote{Notice that in the flat-space limit, the upper bound on
the mass disappears.
This limit cannot therefore rule out the above analysis. It can, however, 
serve as a
guideline to check the consistency of the results obtained in \ads.}.

As was already stressed, the consistency of a string with a finite
number of mass levels is questionable. One can in particular wonder what
the issue of modular invariance could be. Following our discussion of
Section
2, it appears that modular transformations of characters for generic
values of ($k,j$) (see Eq.~(\ref{mt2})) violate the bound (\ref{uc}).
Of course, special values of the spin where singular vectors appear in the
Verma module can lead to characters with different modular properties, and
modular-invariant combinations could eventually be reached. Unfortunately,
interesting situations arise when $2j - k +n=1$ \cite{mff}, which is out
of the would-be unitarity range. Anyhow, since we do not know the
$SU(1,1)$
characters in the
regime $k<-2$, we cannot go any further in the present analysis. 

Finally, the question to be answered is still whether the above
condition (\ref{uc}) can indeed help to restore unitarity. 

As already mentioned earlier, at level $N$, there appears one
representation
of the global algebra with spin $j+N$, which is Virasoro-primary. 
Constraint (\ref{uc}) combined with mass-shell condition (\ref{msc})
is sufficient to guarantee that $j+N$ never becomes positive. Unitarity
also requires the highest-weight vector of that representation to be
positive-norm. This makes condition (\ref{uc}) necessary and sufficient
(the norm vanishes at $k=2j$).

There also appears at level $N$ a Virasoro-primary representation with
spin
$j-N$; since $j-N$ is always negative, all we must check is the norm of
its
highest-weight vector. For $j<-1/2$ and $k<-2$, its norm is strictly
positive, at any $N$ (it vanishes at $j=-1/2$ and is negative for
$-1/2<j<0$, but in this range only level zero is allowed by (\ref{msc})).
Condition (\ref{uc}) plays no role here.

Although technically involved, it is quite straightforward to prove
explicitly the absence of negative-norm states at both level one and
level two
\cite{pmp}. The first level is the only one allowed by condition
(\ref{uc}) for a pure WZW $\su$ model (see Eq.~(\ref{mml}) with
$k=-52/23$), and does not contain, in that case, other Virasoro primaries.
Unitarity is therefore proved\footnote{By mass-shell condition
(\ref{mml}), level two would be allowed for
$j= -1/2 - \sqrt{47/23}/2<k/2=-26/23$. Condition (\ref{uc}) is violated,
and
unitarity is lost.}. 

In order to see what happens at level two, i.e. which
representations survive the Virasoro constraints, we must consider some
extra unitary conformal field theory. It is simple, and quite instructive
as far as counting of states is concerned, to add $d$ free bosons.
The total central charge is now $d+3k/(k+2)$, whereas the space-time
dimension becomes $D=d+3$. The critical dimension is $D_{\rm cr}=29-
{3k\over k+2}$. 

At level one the total number of representations of the global $\su$
algebra is $D$ ($1$~with spin $j+1$, $D-2$ with spin $j$, and $1$ with
spin
$j-1$). Among them, $D-1$ are Virasoro-primary: $1$ with spin $j+1$, $D-3$
with spin $j$, and $1$ with spin $j-1$. On shell, $D-2$  
have positive norm, and $1$ has zero norm (with spin $j$). 

At level two, the total number of representations is $D(D+3)/2$; $1$ has
spin $j+2$, $D-1$ have spin $j+1$, $D(D-1)/2$ have spin $j$, $D-1$ have
spin $j-1$, and $1$ has spin $j-2$. There are $(D+2)(D-1)/2$ Virasoro
primaries: $1$ with spin $j+2$, $D-2$ with spin $j+1$, $(D-1)(D-2)/2$ with
spin $j$, $D-2$ with spin $j-1$, and $1$ with spin $j-2$. The on-shell
positivity properties of these representations are the following:
\nl \noindent (\romannumeral1) For $D<D_{\rm cr}$: $D(D-1)/2$
positive-norm; $D-1$
zero-norm, among which $1$ with spin $j+1$, $D-3$ with spin $j$, and $1$
with spin $j-1$. 
\nl \noindent (\romannumeral2) For $D=D_{\rm cr}$: $(D+1)(D-2)/2$
positive-norm;
$D$ zero-norm, among which $1$ with spin $j+1$, $D-2$ with spin $j$, and
$1$ with spin $j-1$.
\nl \noindent (\romannumeral3) For $D>D_{\rm cr}$: $(D+1)(D-2)/2$
positive-norm; $D-1$ zero-norm, among which $1$ with spin $j+1$, $D-3$
with
spin $j$, and $1$ with spin $j-1$; $1$ negative-norm with spin $j$.
\nl \noindent We thus conclude that {\it all negative-norm states decouple
from the physical spectrum, provided $D\leq D_{\rm cr}$.}
Unitarity is lost otherwise.

These level-two unitarity properties {\it assume condition (\ref{uc})}. If
$j$
becomes smaller than $k/2$, not only does the extremal representation with
spin $j+2$ become non-unitary, but also $D-4$ representations with spin
$j+1$, and $1$ with spin $j$, no matter if we are below, at, or above
the critical dimension. This emphasizes the role played by our
unitarity condition, and gives some credit to the method we have presented
so far, despite the consistency problems that such a bound on the spin
creates at the level of the spectrum. It is even more puzzling that
a real no-ghost theorem might exist, based on the above observations and
more specifically on the constraint (\ref{uc}) over the spin of the
allowed discrete representations. Various works seem to confirm this
viewpoint \cite{hegp}.

As a final remark, I would like to use the above on-shell counting of
states to infer what the partition function would look like, at least for
the contributions originated from the representations of the current
algebra based on the discrete series. String partition functions count
precisely on-shell states -- up to level-matching condition. We thus 
obtain\footnote{Expression (\ref{zexp}), except for the infinite degeneracy 
level by level, is
similar to the corresponding one for the free bosonic string. This is due to 
the
structure of $\su$ algebra, and does not hold at higher levels.}:
\ba
Z(\tau, \bar \tau, v, \bar v) \sim  v^{-1} q^{-1} \Bigg(1    &+& 
\left( \e^{2i\pi v} +D-4 +\e^{-2i\pi v} \right) q    \nn \\    &+&
\Bigg( \e^{4i\pi v} + \e^{-4i\pi v} + 
       \left( \e^{2i\pi v} +\e^{-2i\pi v}\right) (D-3) \nn \\    &&
       + {1\over 2}(D-2)(D-3)             \Bigg) q^2 + 
       {\rm O}\left(q^3\right) \Bigg) \times {\rm \ c.\  c.} \label{zexp}
\ea
(here $26>D\ge 3$ is a free integer parameter, and the level 
$-\infty < k \leq -52/23$ is chosen
such that $D=D_{\rm cr}$). 
Expression (\ref{zexp}) can be seen as the 
expansion of 
\be
Z(\tau, \bar \tau, v, \bar v)= { F(\tau, \bar \tau, v, \bar v)
\over 
(\im)^{D-5\over 2} \, \vartheta_1 (\tau,v)\,  \bar \vartheta_1(\bar \tau,
\bar v) \,
\big(\eta (\tau)\, \bar\eta(\bar\tau)\big)^{D-5} }\; ,
\label{z}
\ee
where $\left(\sqrt{\im} \, \eta \bar\eta\right)^{D-5} $ 
stands for the
free-boson-and-ghost contributions 
and $F(\tau, \bar \tau, v, \bar v)$ 
behaves like
\be
F(\tau, \bar \tau, v, \bar v) = \e^{-{\pi \im \over  k+2 }}\, \e^{-2\pi
\Im v} 
\left(1 +{\rm O}\left(q^3\right) \right) \left(1 +{\rm O}\left({\bar
q}^3\right) \right)\; .
\label{fexp}
\ee
In fact, any power of $q$ and $\bar q$ ($\ge 3$) is expected
as a consequence of modular covariance:
$$
F\left(-{1 \over \tau}, -{1 \over \bar \tau}, {v \over \tau}, {\bar v  
\over
\bar \tau}\right)=
\left\vert \e^{-2i\pi {v^2 \over \tau}}\right\vert
\left\vert \tau\right\vert
F(\tau, \bar \tau, v, \bar v) \; ,
$$ 
whereas
$
F(\tau, \bar \tau, v, \bar v)$ should be invariant under $\tau \to \tau
+1$. These constraints can actually  be
satisfied with expressions that do not fall in the class of (\ref{fexp}),
such as $F=(2 \im )^{-1/2} \exp  2 \pi {(\Im v)^2 \over
\im }$, inspired from Eq. (\ref{dni}). The latter expression for $F$
is actually what (\ref{dni}) would have given if the measure 
${\rm d}j$ had been replaced with ${\rm d}j$ $\exp - \pi k{(\Im
v)^2\over\im}$. I have already discussed this issue in Section 2 for the
pure WZW model. Here, it becomes clear that 
such a function is not allowed since {\it it does not exhibit the correct
weight shift to fulfil the mass-shell condition.} 
Furthermore, at large   
$\vert k \vert$ ($D \to 26$) and small $v$, matching with
the ordinary $26$-dimensional string requires the following 
behaviour\footnote{This is precisely the next-to-leading behaviour of
expression (\ref{dni}). The leading term diverges like $\sqrt{k}/\vert
v\vert^2$, and should be avoided in the presence of world-sheet
supersymmetry. The various uncertainties related to the meaning of
(\ref{dni}), however, do not
enable us to draw any conclusion.}:
$F\to \kappa \vert v \vert^2 (\im)^{-3/2},$ where $\kappa$ is a constant
expected to diverge like $\vert k \vert^{3/2}$.
In order to obtain correctly the various behaviours, one should
therefore
rely on the generic form  
(see Eq.~(\ref{zgen}))
$F(\tau, \bar \tau, v, \bar v)=
\sum_{j,\ell} {\cal N}_{j,\ell}\,
f_{j}(\tau,v)\,
\bar f_{\ell}(\bar \tau, \bar v)
$, where $f_{j}(\tau,v)$ is the factor in the character
of the spin-$j$ representation, which
accounts for the null states in the Verma module.
The presence of terms of ${\rm O}\left(q^3\right)$  
and
${\rm
O}\left({\bar   
q}^3\right)$ precisely traces back the appearance of representations of
the $\su$
current algebra in the spectrum, which contain {\it  null states at levels 
higher than two.} As was already mentioned before, such
representations are
expected to appear for $j\leq k/2$; this bound is in contradiction with
the unitarity constraint (\ref{uc}).

Obviously, 
using naive generic-($k,j$)
characters (\ref{chg}) {\it cannot} lead to an expression for $F(\tau,
\bar
\tau, v, \bar v)$ consistent 
with (\ref{fexp}). Furthermore, it is hard to believe that $F$ exists,
such that expression (\ref{z}) is polynomial in $q, \bar q$ of degree 
$N_{\rm max}$, as is suggested by the previous study of unitarity.
Once
more, {\it compatibility between unitarity and modular invariance} appears
as an
important issue in understanding the string propagating over \ads.

\boldmath
\section{Summary and comments}
\unboldmath  

String theory on three-dimensional anti-de Sitter space-time has been
commented for a long time. No satisfactory understanding of its basic
features, such as the complete spectrum of perturbative states, has yet
been reached. By formally using the standard current-algebra approach of
conformal field theory in the framework of a non-compact group manifold,
it seems that unitarity would demand an upper bound to the mass spectrum.
This conclusion is in disagreement with elementary principles of string
theory, leads to serious problems in computing one-loop amplitudes, and
has no physical foundation: nothing similar appears in the 
quantum motion
of a free particle on \ads, which is expected to be the $\alpha ' \to 0$ 
limit of the string, and where {\it all} possible representations of $\su$
appear in the wave-function, with appropriate interpretation \cite{bs}. 

Nevertheless,
it is quite amazing that the cut-off over the spins could really fix the
unitarity problem \cite{pmp, moh, hegp}, and this may be a good reason
to try to keep 
the above results, and rephrase them within a somewhat less formal
approach. This means that we should get a better understanding of the
$\su$ WZW model at the {\it classical level}\footnote{See \cite{banl}. In
that
spirit, $\su /U(1)$ has been recently revisited \cite{mw}. Remember that 
$\su /U(1)$ was analysed in \cite{dlp} as an internal conformal field
theory for a string compactification. Unitarity was proved, provided 
$k/2\leq j <0$. Later, $\su /U(1)$ was reinterpreted as a two-dimensional
black hole \cite{2dbh}, and the spectrum was studied in \cite{dvv}. This
could also be a source of inspiration for $\su $ itself, 
by considering $\left(\su / U(1) \right) \times U(1)$. There
are many ways to define the latter, with various geometrical
interpretations -- when available \cite{bk,koun}.}, explore the 
classical motion of a string,
and maybe try 
to follow a path-integral approach {\it \`a la}  Gaw\c{e}dzki~\cite{gaw}.
This
implies in particular a proper treatment of the {\it target-space boundary
conditions}
that are hard to implement within the current-algebra method.
They 
might play a role in reconciling unitarity with the appearance
of  states of the discrete series with spin $j<k/2$, i.e. 
with mass above the anti-de Sitter radius scale.
These states are cut off in our analysis but
should
be present for physical reasons, and might originate from some {\it other
unitary sector} of the theory, which would have been missed 
here\footnote{A similar viewpoint was somehow taken by the authors
of \cite{mod}. Their subsequent developments were, however, quite ad-hoc,
and the net result for the partition function looks more like an analytic
continuation of an $SU(2)_k$ invariant than like a true amplitude 
computed from first principles in the theory under consideration.}. The role 
of continuous series could also be
clarified. Although they are compatible with unitarity, the corresponding
excitations seem to be all tachyonic. Finally, one should wonder whether  
target-space boundary conditions introduce  {\it ambiguities in the
quantization
of a string}, similar to those appearing when solving the wave equation
for
a quantum particle \cite{oldads3}. Such ambiguities may have interesting
consequences for the string, as they have when studying e.g. the Unruh
effect on \ads\ \cite{deser}.

As was pointed out previously, one could try alternatively to avoid
the mass/spin cut-off and the purely tachyonic continuous
representations in various ways, playing essentially with the
current algebra, and/or modifying the affine Sugawara
construction. It even seems that contact with some ADS/CFT-inspired
results can be made in that way \cite{barsn}. However, it is not clear
whether such modifications leave unaltered the interpretation of the
theory as a string propagating over \ads\ 
(in \cite{bars}, logarithmic cuts and a new zero mode are introduced in
the currents, and representations based on discrete series are simply
discarded). 
Furthermore, none of the
available attempts treats the {\it problem of the infinite degeneracy}
at each string level, or enlightens the issue of modular invariance.
In particular, the role of the ``magnetic field" $v$ remains obscure. The
possibility of interpreting it as a continuous twist, similar to those
that appear in the parafermionic constructions, has never been exploited.
One might, though, relax in this way the constraint of modular
invariance:
the latter should be recovered only after an appropriate summation over
$v$ (like in the case of the two-dimensional Euclidean black hole
\cite{gaw}).

Finally, a somewhat more exotic attitude\footnote{An even more marginal
alternative would be to drop the requirement of modular invariance, and
simply face the ``stringy exclusion principle" \cite{mast}. It is hard
to believe that this could be the end of the story, since so many features
of string theory, such as the IR/UV duality, strongly rely on modular
invariance.} with respect to the
unitarity
could be to simply admit the presence of negative-norm states in the
physical spectrum, and then try to interpret them or, better, to identify
the
instability they are related to and its physical origin
within the \ads\ background. Following this line of thought, one could
even reconsider the -- non-unitary -- models based on the admissible
representations of the $\su$ current algebra; their partition function is
known, and one should then try to understand them in the framework of
string theory.

Let me emphasize once more that the motivations for
studying the string on \ads\ are wider than expected in the early works.
They
include the string motion in a three-dimensional black hole \cite{sat},
which settles the proper framework to address the black-hole evaporation
problem.
Instead of a NS--NS torsion background, coupling to non-perturbative R--R
charges is also a relevant and difficult problem
\cite{rr}. Finally, the analysis of the
ADS/CFT 
conjecture, in the framework of the
\ads $\times S^3$ background\footnote{Notice that \ads $\times S^3$
modded          
out by some discrete symmetry might be more tractable than \ads $\times
S^3$ itself, in particular when we demand supersymmetry.},
is probably the issue that has attracted 
most attention. Ideally, we would like to compute correlators in both
sides and compare them. In practice, correlators for \ads\ string states
are out of reach, which
makes any rigorous check quite
intricate. Therefore, most of the
work in that direction has been devoted to  trying to express the
space-time as well as the
asymptotic two-dimensional conformal symmetry, in terms of the fields
of the WZW model 
whose target space is the bulk \ads\ theory, and to build in
that
way the boundary conformal field theory. It is fair to say, however, that 
this approach has not shed any light on the structure of the \ads\
string itself
-- at least regarding the questions raised here; 
as long as one does not handle the \ads\ side exactly, the
achievements
are limited both on checking the ADS/CFT correspondence, and on building
the boundary theory \cite{boundconf}. Of course, there is still the
-- weaker -- alternative 
to work with the low-energy supergravity, supplemented with all
Kaluza--Klein excitations coming from higher dimensions, thus trying to
obtain some feedback for the string on \ads.
For example, there are signs that 
all $\su$ representations -- discrete and continuous -- should appear
without bound on the mass. If such a bound were present, 
assuming the ADS/CFT correspondence,
it
would be hard to identify states 
in the bulk \ads\ supergravity (or string, as a fundamental theory), with
states in the boundary conformal field theory. 

As a last comment, I would like to stress that string theories on more
general ADS$_n$ backgrounds are equally 
important and more difficult than the three-dimensional case at hand.   
In odd dimensions, it has been realized very recently
that both an antisymmetric tensor and a linear dilaton are needed
together with the gravitational background, in order to define an exactly 
conformal sigma model describing the string \cite{dbs}. This sigma model
{\it is not}
a WZW model, and its spectrum and interactions remain
quite unexplored. Furthermore,
the exact conformal field theories describing the string
propagation on even-dimensional anti-de Sitter spaces
have never been investigated. 
Notice also that some issues, such as the 
level-by-level infinite degeneracy, or even the unitarity problem, will be
generically present.

\vskip 1.cm
\centerline{\bf Acknowledgements and apologies}
\noindent
It is a pleasure to thank C. Bachas and P. Bain for the stimulating
collaboration we have been having on this subject. I would also like
to
acknowledge many colleagues, from the CPT in Ecole Polytechnique and the
LPT
in Ecole Normale Sup\'erieure, for interesting discussions.
The material I presented here is not to be considered as a review
over the very wide subject that \ads\ has become on its own. Many
interesting aspects are missing or superficially mentioned.
I hope that the reader will forgive me for the very incomplete list of
references. 
Finally, I would like to thank the organizers of the XXXIst Summer
Institute held in ENS, Paris, 16--31 August 1999, and of the 
TMR (ERBFMRX-CT96-0045) European program meeting ``Quantum aspects of
gauge theories,
supersymmetry and unification", Paris, 1--7 September 1999. 
This work was  supported in part by EEC TMR contract
ERBFMRX-CT96-0090.

\end{document}